# Fast non-thermal switching between macroscopic charge-ordered quantum states induced by charge injection.

I. Vaskivskyi[1], I. A. Mihailovic[1], S. Brazovskii[2], J. Gospodaric[1], T. Mertelj[1], D. Svetin[1], P. Sutar[1] and D. Mihailovic[1]

[1] *Jozef Stefan Institute, Jamova 39, SI-1000 Ljubljana, Slovenia*
[2] *LPTMS-CNRS, UMR8626, Université Paris-Sud, F-91405 Orsay, France.*

**The functionality of logic and memory elements in current electronics is based on multi-stability, driven either by manipulating local concentrations of electrons in transistors, or by switching between equivalent states of a material with a degenerate ground state in magnetic or ferroelectric materials. Another possibility is offered by phase transitions with switching between metallic and insulating phases, but classical phase transitions are limited in speed by slow nucleation, proliferation of domains and hysteresis. We can in principle avoid these problems by using quantum states for switching, but microscopic systems suffer from decoherence which prohibits their use in everyday devices. Macroscopic quantum states, such as the superconducting ground state have the advantage that on a fundamental level they do not suffer from decoherence plaguing microscopic systems. Here we demonstrate for the first time ultrafast non-thermal switching between different metastable electronically ordered states by pulsed electrical charge injection. The macroscopic nature of the many-body quantum states(*1-4*) - which are not part of the equilibrium phase diagram - gives rise to unprecedented stability and remarkably sharp switching thresholds. Fast sub-50 ps switching, large associated resistance changes, 2-terminal operation and demonstrable high fidelity of bi-stability control suggest new opportunities for the use of macroscopic quantum states in electronics, particularly for an ultrafast non-volatile quantum charge-order resistive random access memory (QCOR-RAM).**

Competing interactions between electrons in materials exhibiting macroscopic charge order may support a variety of ground states at different temperatures or pressures, leading to complex phase diagrams. The layered dichalcogenide *1T*-TaS$_2$ is a particularly important system in which competition between Coulomb interactions, lattice strain and a

Fermi surface instability(*5-10*) lead to a number of macroscopic quantum states with different electronic charge order: a metallic state with an incommensurate charge density wave (CDW), a domain textured nearly-commensurate (NC)(*11*) or an insulating commensurate (C) state (*12*), and even superconductivity(*13*). At room temperature, the material is in the NC state, which can be described as a periodic texture of ordered polaron clusters, each polaron being in the form of a hexagram (Fig. 1 a). On cooling, it undergoes a first order transition to the C state near 170K, accompanied with an abrupt change of resistance R and insulating behaviour at low T (*12*) as shown in Fig. 1b. The configurations of electronic ordering corresponding to the NC and C minima in the equilibrium free energy landscape of *1T*-TaS$_2$ (*8, 9*) are shown schematically in the insert to Fig. 1a. Here we show that under highly non-equilibrium conditions created by pulsed charge injection, multiple metastable states can be reached which are not part of the equilibrium phase diagram. The remarkable stability of the new state, whose extrapolated lifetime exceeds the age of universe below ~20 K is highly unusual. The stability can be influenced by strain and is modified upon reducing the dimensionality.

Thus, a single pulse from a constant current source (2 mA at 10V, 1 μs) applied to *1T*-TaS$_2$ at 4 K (see Methods for details) leads to an immediate drop in R by nearly three orders of magnitude (Fig. 1 b). Measuring R upon heating, it gradually merges with the cooling curve at around $T_H$ ~ 80 K (Fig. 1 b). To investigate this striking phenomenon further, we first incrementally increase the current under *constant current* conditions and measure the voltage during 50 μs current "write" (W) pulses (The circuit is shown in Fig. 1c). The resulting I-V curve (Fig. 1d) shows a number of remarkable features. Up to ~ 2 V, the I-V curve is linear (Fig. 1 e). At approximately 2 V, there is a small but sharp discontinuity, where at $I_{c1}$ = 0.21 mA the voltage drops by ~ 0.1 V. (Fig. 1 f) Thereafter (> 3V) the current increases *exponentially* with V, up to $V_T$ = 8.1 V, which can be described quite well by I=$I_0$ exp(V/$V_0$) with fitted values of $V_0$ = 7.1V and $I_0 \approx$ 0.02 mA (Fig. 1 g). The observed linear-to-exponential I-V characteristic with large $V_T$ in Fig. 1 b and g is not compatible with simple microscopic electronic transport models, and its functional form is very different than for a sliding CDWs(*14-16*). At a threshold $V_T$ ~ 8.1 V (and ~ 2mA for this device), the voltage drops within a remarkably narrow current interval $\Delta I$ < 40 μA (Fig. 1h), and the I-V curve becomes linear (ohmic), extrapolating to the origin. The system remains in this low resistance state until heated above $T_H$.

Applying a longer (10 s) electrical pulse to raise the temperature of the device by Joule heating leads to sharp switching of the resistance back to a high resistivity state. In the first cycle the resistance value reached after such an "erase" (E) process is slightly lower than the original one. In subsequent cycles the resistance stabilizes, varying by ~ 10 % around a value which is ~ 30% below the original one (Fig. 2 a). It also fluctuates in time by ~ 1 Ohm when measured over longer periods (Fig. 2 b). In contrast, the low resistance state reached after the W pulse stable to < 0.03 Ohms over long periods (hours) provided the device temperature is below ~ 40 K (Fig. 2 c).

Applying a series of W pulses, incrementally increasing the voltage under *constant voltage* conditions, we find that several intermediate resistance states are reached (Fig. 2 d). In the states reached following 3.2 and 6.4 V pulses, the resistance relaxes slightly in ~ 1000 s (see insert to Fig. 2 d). Remarkably, for V = 16 V there is no such relaxation, and the resistance remains constant after the W pulse. The different states thus appear to have different relaxation properties.

Measuring the threshold voltage $V_T$ as a function of distance between contacts, we find a linear dependence (Fig. 2 e). $V_T$ is always higher for the first W pulse, and settles thereafter to a value which is ~ 20% lower. The threshold electric field $E_T = 1.7 ~ 2.5 \times 10^4$ V/cm is much higher than the 10 V/cm nonlinearity threshold previously reported for the C state (*17*), and is closer to the sliding threshold in other related materials, such as $NbSe_3$ (*18*). $I_{c1}$ and $V_0$ were not found to depend on L.

In Fig. 3 we show the switching of the device for different pulse length $\tau_W$ using W/E sequences. Switching is observed for all pulses with $\tau_W <$ 1 s to ~ 30 ps (limited in speed by the current generator circuit). For $\tau_W >$ 1 s, approaching the previously mentioned E pulse properties, switching is no longer observed. The abrupt E threshold strongly suggests that a critical temperature needs to be reached for the system to revert to the C state. Bi-stable switching is demonstrably achievable also with intermediate states. In Fig. 3 b we show the pulse length dependence of switching of an intermediate state with TTL-level 4V pulses with $\tau_W = 10^{-5}$.

The temperature dependence of the I-V curves is shown in Fig. 3 c. Remarkably, we see that switching takes place up to 200 K, with two distinct temperature regions (up to ~ 100K, and between 150 ~ 200 K) where the resistance switching ratio is more pronounced. Overall the resistance jump becomes smaller at higher T. Remarkably, $I_T$ is T-independent. The temporal relaxation of the low-resistance state is shown for various temperatures in Fig. 3 d. An Arrhenius plot of its lifetime from fits (Fig. 3d) to the relaxation data shown in Fig. 3 e gives extrapolated lifetimes of ~ 2 x $10^{54}$ s at 4 K, and ~ 0.2 s at 200 K. This indicates that the $T_H$ shown in Fig. 1 b is not intrinsic, but is determined by the measurement time.

Resistance switching by electrical pulses in other typical systems is either of a percolative(*19, 20*) or filamentary nature(*21*), or arises as a result of a structural phase transition(*22, 23*). The switching observed here is most of all reminiscent of recently observed switching in superconducting systems(*24*).

The CDW behaves as a deformable quantum solid, whose wavevector *q* depends on the charge density. Transforming the system from one state to another requires a macroscopic change of *q*, subject to boundary conditions imposed by the contacts. The C state (Fig. 1a) has a periodicity of $2\pi/q_c = \sqrt{13} \times \sqrt{13}$ lattice constants, rotated with respect to the underlying lattice basis vectors by 13.9 degrees. The *q* of an incommensurate CDW, relative to $q_c$ at any particular point between the contacts is directly related to the excess charge density by $(q-q_c)/\pi = \delta n$, where $\delta n = n_e - n_h$ is the local charge imbalance (*25-27*). After the injection of electrons and holes at the two contacts respectively, the electrons and holes propagate along the sample (Fig. 3 f-h). If the mobilities of the two types of charge carriers are not equal, such as when the electronic structure is highly asymmetric with respect to the Fermi level as in the C state of *1T*-TaS$_2$ (*28*), then after equilibration we can imagine that before e and h completely recombine, an incommensurate electronically ordered state can form with δn ≠ 0 and q ≠ $q_c$ (Fig. 3 h). This leads to an additional conductivity channel, whose resistance is determined by the density of carriers via $n_c$ ~ (*q*-$q_c$)/π. The extremely sharp switching thresholds are nevertheless surprising, indicating that a collective many-body ordering process is responsible for both W and E processes, with quantum coherence over distances of at least a few microns (the distance between contacts). We can speculate that apart from different charge ordering within individual layers, layers may also exhibit a variety of stacking configurations, which are quite close in energy(*29*).

Once formed, the transformation back from one *q* to another requires an energetically costly topological transformation: as a result of macroscopic quantum coherence, the addition of each extra wave in the CDW between the contacts requires not only a charge rearrangement of the entire macroscopic system, but also a charge carrier conversion from the underlying fermionic reservoir provided by the normal electrons (which are not involved in the CDW) to electrons in the CDW(*30*). Such a quantum "protection" mechanism can be very effective in stabilizing the different states. Of course it is expected that pinning by defects, substrate strain and other external factors will have an additional role in the dynamics of the transformation.

The very fast non-thermal switching, sharp threshold behavior in both W and E and simple device operation promises a new generation of ultrafast quantum charge-order resistive random access memory (QCOR-RAM) devices. The electrical W switching speed of ~ 30 ps shown here is not limited by the intrinsic mechanism but by the rise time of the pulse source, yet it is ~ 3 orders of magnitude faster than state of the art PCM devices(*23*) or memristors(*31*). With appropriate circuits and THz sources, we may expect to achieve electrical switching on sub-picosecond timescales observed optically(*30*). Here we demonstrated switching up to 200K, but if room temperature operation is considered, we note that unlike with superconductors, CDW systems with competing ground states already exist above room temperature, so it's reasonable to expect that materials with suitable energy landscapes supporting switching between different CDW states can be found in the foreseeable future.

**Methods summary.**
Transport-grown single crystal samples between 20 and 120 nm thick, were deposited on sapphire substrates by exfoliating them with sticky tape and re-depositing onto the substrate. Multiple gold contacts were deposited using laser direct photolithography with a LPKF Protolaser LDI laser writer. A 5 nm Au/Pd intermediate layer beneath 100 nm thick gold electrodes was deposited by sputtering. The resistance is measured either in 2 or 4 contact configuration, unless otherwise stated, with low currents < 1 μA to minimize perturbation of the system by the measuring current. The switching was performed with electrical pulses using a Keithley measurement system (Keithley 6221 Current source + Keithley 2182A Nanovoltmeter) for "write" (W) pulse duration $\tau_W$ > 5 μs. For intermediate pulse durations 20 ns < $\tau_W$ < 5000 ns we used a Stanford DG535 pulse generator, and an TI THS4211EVM amplifier. For pulses $\tau_W$ < 20 ns, we used a Hamamatsu MSM G4176-03 photodetector with a pulsed femtosecond laser source generating current pulses with a 30 ps rise/fall time. The experiments were performed in a Lake Shore 4-probe measuring station, a He-flow cryostat or a closed cycle Oxford instruments cryostat with semi-rigid RF cable connections made directly to the sample (for the ultrafast switching experiments). The measuring circuit and a typical sample showing multiple contacts used for measurements is shown in Fig. 1c.

**Acknowledgments**.
DM wishes to acknowledge funding from the Advanced ERC grant "Trajectory".

Figure 1. **Resistance switching behavior of *1T*-TaS$_2$ resulting from the application of short electrical pulses. a,** A schematic free energy landscape of *1T*-TaS$_2$ as a function of wavevector $q$ (after (*9*)). q$_I$, q$_{NC}$ and q$_C$ are the wavevectors corresponding to the incommensurate (I), nearly commensurate (NC) and commensurate (C) states respectively, shown schematically in real space. The individual polaron is also shown. **b,** The resistance on cooling from room temperature to 4K (blue symbols). *After* an electrical W pulse (2 mA, 10 V, 1 µs) is applied at 4 K, the resistance drops to 80 Ohms. The sample is then heated to 100 K (red symbols). The arrow indicates the change of resistance after a W pulse is applied. **c,** The measuring circuit with a typical sample and contacts. Here, the sample is 130 nm thick and the gap between contacts is L=4.5 µm. **d**, An I-V curve measured at 14 K in pulsed mode ($\tau_W$=50 µs), where the current is incrementally increasing, and the voltage across the sample is measured simultaneously. The main features of the I-V curve are shown expanded: **e**, a linear I-V characteristic up to 2 V. **f,** a sharp discontinuity at ~ 2 V, where after the current increases exponentially with voltage as shown in the log-linear plot **g**. **h**, the switching occurs in a narrow current interval, at 8.1V for this sample.

Figure 2. **The resistance after repeated cycling. a)** The final state reached after the W pulses appears to always have the same resistance, to within < 2% (here V = 20 V, $\tau_W$ = 5 µs, T = 26 K), while the resistance after the E pulse (V = 7 V, $\tau_E$ = 10s) varies somewhat (note the logarithmic scale). **b** the relaxation of the resistance after an E pulse and **c)** the resistance after the W pulse on the same scale for a typical cycle (indicted by blue and red circles in **a**. **d** a resistance cascade through intermediate states obtained by gradually increasing the switching voltage near threshold ($\tau_W$ = 2 µs) for a device with a 3 µm gap between contacts at T = 4 K (note the logarithmic scale). The insert shows the relaxation of the intermediate states after 3.2 and 6.4 V pulses on a linear scale. **e** The dependence of $V_T$ on the distance between contacts L . $V_T$ is slightly higher in the first cycle than in subsequent cycles (see also **a**).

Figure 3 **The dependence of the switching on the pulse length $\tau_W$ and temperature.
a**. The resistance of the sample is shown after the application of 16 V pulses of different $\tau_W$ at 26 K. **b** switching with transistor logic level pulses (4V) to an intermediate state. **c**. T-dependence of I-V curves. Switching is no longer detected above 200 K. The arrows 1-6 show the measurement cycle, which is repeated for each temperature. **d** Relaxation of the switched state after switching at different temperatures. **e** An Arrhenius plot of the relaxation time of the switched state. **f-h**, The spatial variation of the CDW modulation between the contacts at different times. The blue line depicts the spatial variation of the charge density ρ, the red line depicts the spatial variation of δq. Electrons (e) and holes (h) are injected (**g**) at the electrodes initially in the C state (**f**). They then move through the crystal at different rates (**g**). Asymmetry of the carrier transport leads to a state with δq≠0 as shown in **h**.

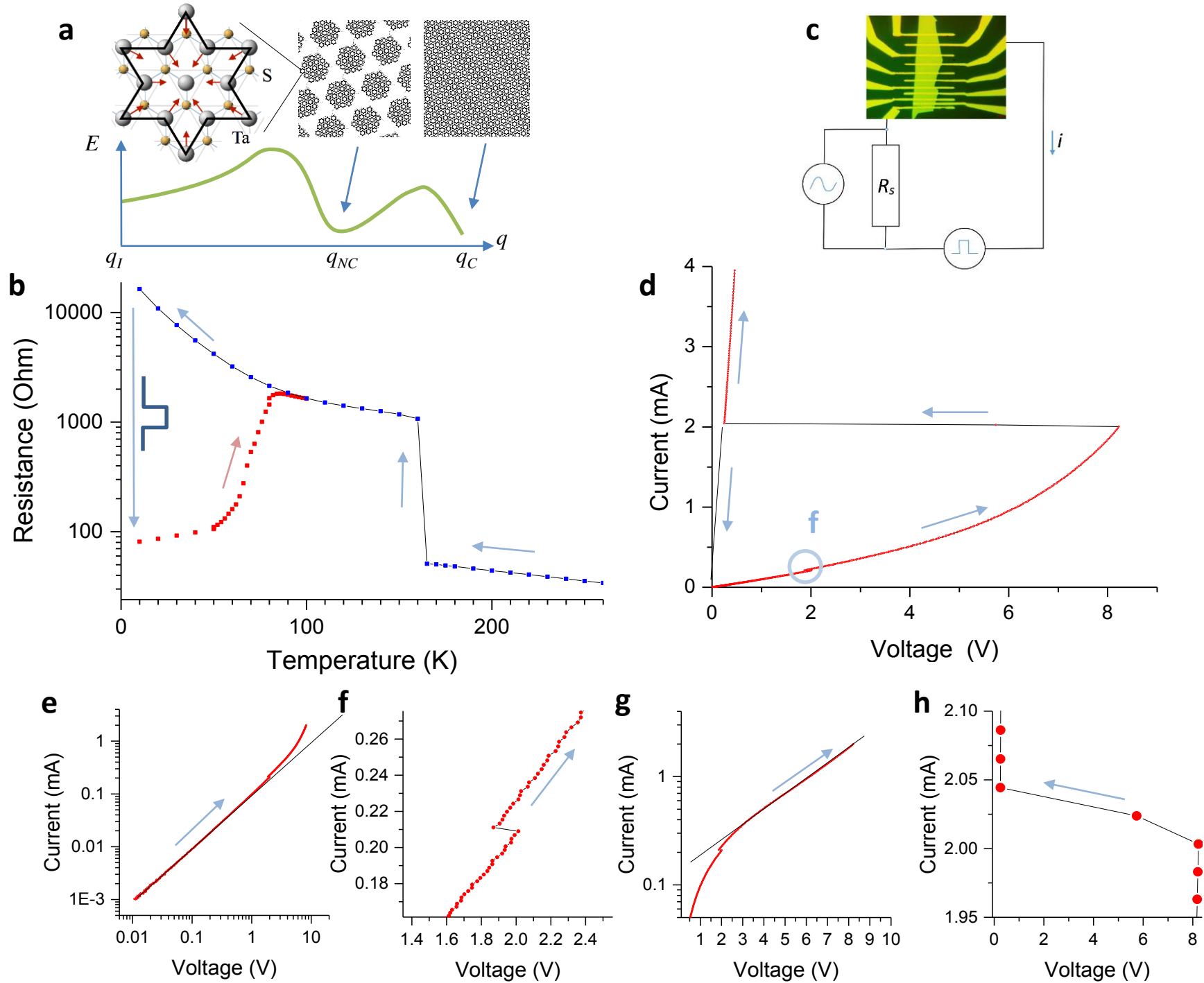

Fig. 1

Fig. 2

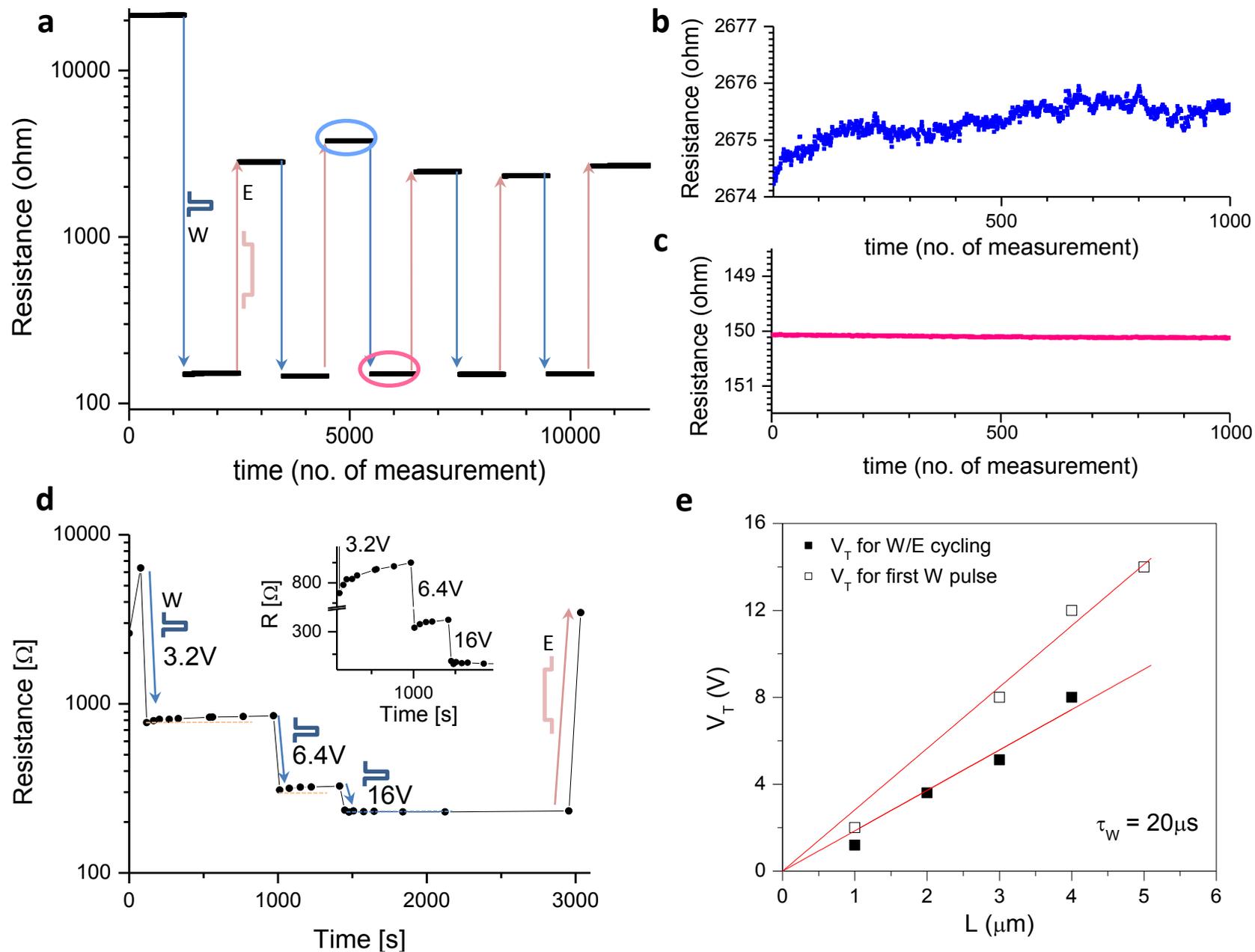

Fig. 3

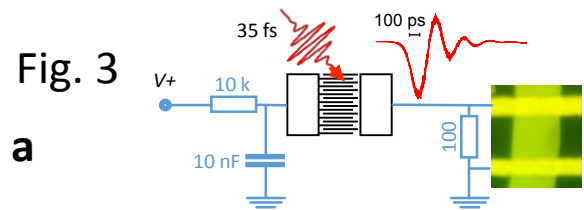
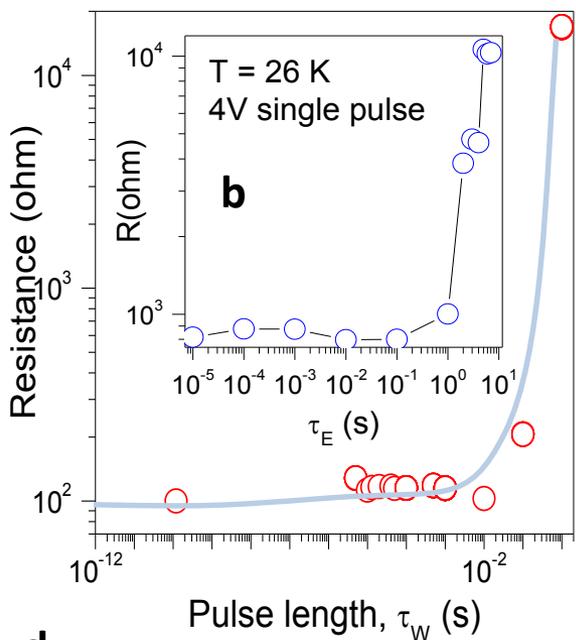
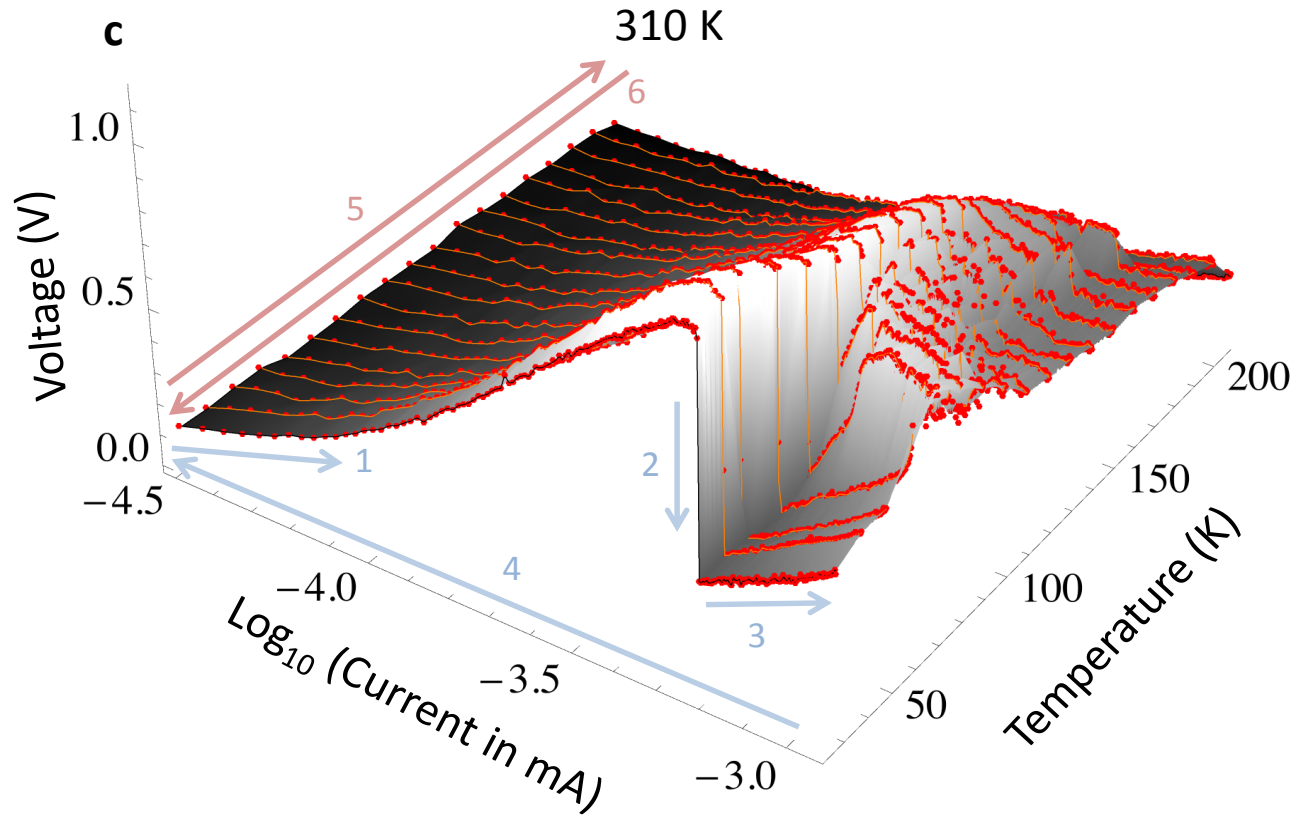
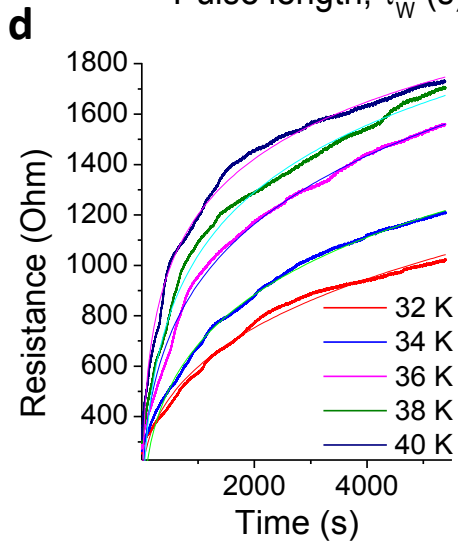
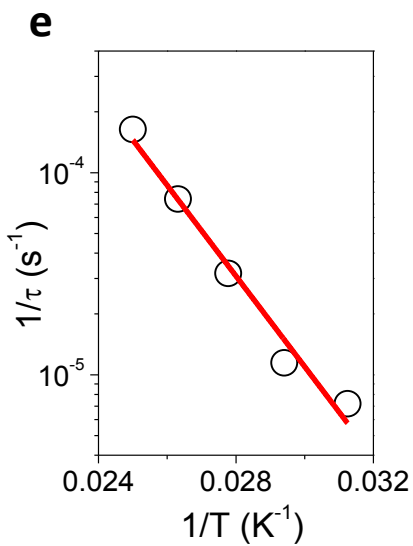
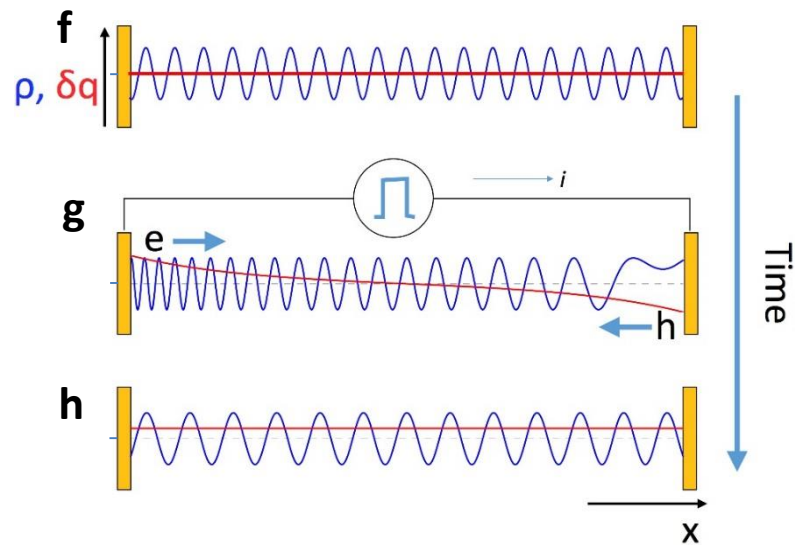